\documentclass[twocolumn]{aastex63}

\usepackage[colorinlistoftodos]{todonotes}
\let\Oldtodo\todo
\renewcommand{\todo}[1]{\Oldtodo[inline]{#1}}

\usepackage{amsmath}
\usepackage{float}

\shorttitle{Migration $\&$ the Giant Planet Occurrence Rate}
\shortauthors{Hallatt \& Lee}

\graphicspath{{./}{figures/}}

\begin{document}

\title{Can Large-Scale Migration Explain the Giant Planet Occurrence Rate?}

\correspondingauthor{Tim Hallatt}
\email{thallatt@physics.mcgill.ca}

\author[0000-0003-4992-8427]{Tim Hallatt}
\affil{Department of Physics and McGill Space Institute, McGill University, Montr\'eal, Qu\'ebec, H3A 2T8, Canada}
\affil{Institute for Research on Exoplanets (iREx), Montr\'eal, Qu\'ebec, Canada}

\author[0000-0002-1228-9820]{Eve J.~Lee}
\affil{Department of Physics and McGill Space Institute, McGill University, Montr\'eal, Qu\'ebec, H3A 2T8, Canada}
\affil{Institute for Research on Exoplanets (iREx), Montr\'eal, Qu\'ebec, Canada}

\begin{abstract}

The giant planet occurrence rate rises with orbital period out to at least $\sim$300 days. Large-scale planetary migration through the disk has long been suspected to be the physical origin of this feature, as the timescale of standard Type I migration in a standard solar nebula is longer farther from the star. These calculations also find that typical Jupiter-bearing cores shuttle towards the disk inner edge on timescales orders of magnitude shorter than the gas disk lifetime. The presence of gas giants at myriad distances requires mechanisms to slow large-scale migration. We revisit the migration paradigm by deriving model occurrence rate profiles from migration of cores, mass growth by gas accretion, and planetary gap opening. We show explicitly that the former two processes occur in tandem. Radial transport of planets can slow down significantly once deep gaps are carved out by their interaction with disk gas. Disks are more easily perturbed closer to the star, so accounting for gap opening flattens the final orbital period distribution. To recover the observed rise in occurrence rate, gas giants need to be more massive farther out, which is naturally achieved if their envelopes are dust-free. The range of mass gradients we find to reconcile the observed occurrence rate of Jupiters is too abrupt to account for the mass-period distribution of low-eccentricity giant planets, challenging disk migration as the dominant origin channel of hot and warm Jupiters. Future efforts in characterizing the unbiased mass distribution will place stronger constraints on predictions from migration theory.

\end{abstract}


\section{Introduction} \label{sec:intro}

Unlike our solar system, exo-Jupiters appear at a variety of orbital periods, as close as a couple of days to beyond 1000 days. While they are a relatively rare population---with only $\sim$10\% of solar-type stars harboring a giant planet out to 2000 days \citep{cumbutmar08}---the existence of hot and warm Jupiters have challenged our understanding of planet formation, and their origin remains elusive.

Giant planet formation theories currently fall under three classifications: in-situ formation, disk migration, and high-eccentricity migration. 
The high-eccentricity migration paradigm transports giant planets to small periods in a two-step process: first by exciting Jupiters' eccentricities, then by tidal circularization and shrinkage of their orbits due to close passage to their host star. Eccentricities may be excited 
through Kozai-Lidov oscillation induced by a stellar companion \citep[e.g.,][]{wumur03,fabtre07,naofarras12}, by an inclined planetary companion \citep[e.g.,][]{Naoz11,dawchi14}, or by a coplanar but eccentric planetary companion \citep[e.g.,][]{Li14,pet15b}. Planet-planet scattering through dynamical instability in multi-planetary systems may also contribute to the production of hot Jupiters 
\citep[e.g.,][]{rasfor96,nagida11,beanes12}.
Warm Jupiters prove challenging to be produced by high-eccentricity migration. In general, the theory underpredicts the warm to hot Jupiter number fraction \citep[e.g.,][]{pettre16,Hamers17}. Observationally, approximately half of warm Jupiters are accompanied by small, nearby planets \citep{huawutri16} which are difficult to maintain through violent dynamical migration. Furthermore, for select systems with warm Jupiters and an outer planetary companion, \citet{anthamlit16} demonstrate the systems would have been unstable had the warm Jupiter formed beyond 1 AU and underwent high-eccentricity migration.

In disk migration, tidal torque from the surrounding disk gas can shrink a planet's orbital radius from AU-scales to the inner disk edge. Migration has long been recognized as a natural and inevitable consequence of planetary tidal interaction with a disk \citep[e.g.,][]{linpap79,goltre79,goltre80,linpap86}. A longstanding issue with migration is that in a standard solar nebula, the migration timescale is shorter than the disk lifetime by orders of magnitude so additional physics is required to explain the existence of planets well beyond the inner edge of protoplanetary disks \citep[see][for a review]{klenel12}.
Previous studies invoked initial locations that are sufficiently far away \citep[e.g.,][]{armlivlub02,alidib17}, well-timed photoevaporation of the disk carving out a gap at $\sim$1 AU \citep[e.g.,][]{alepas12}, or radial substructures in the disk that act as speed bumps \citep[e.g.,][]{colnel16}.

In-situ formation posits that giant planets are born where we observe them. The often-quoted challenges in creating hot and warm Jupiters lie in avoiding large-scale migration through the disk, and in the disks' ability to breed massive enough cores close enough to the central star. As long as such cores are emplaced, runaway gas accretion can proceed to blow them up into giants \citep{Rafikov06,bolgragla16,batbod16}. Whether a given disk can build requisite core masses in the inner disk depends sensitively on the unknown details of disk structures including the local dust-to-gas mass ratio, dust particle size distribution, and the strength of turbulent ``viscosity'' \citep[e.g.,][]{Lambrechts14,funlee18,Lin18}. Large-scale migration can be avoided if planets form in inviscid disks; without ``viscous'' diffusion to fill planetary gaps back in, enhanced outward feedback torques slow and eventually halt migration \citep[e.g.][]{Li09,Yu10,funchi17}. We emphasize that by in-situ formation, we imply the final assembly of planets that occur largely where they are presently observed. In-situ formation therefore does not preclude initial periods of migration of e.g., planetary protocores, but does point against migration setting the final properties of planets.

The period distribution of hot and warm Jupiters has been refined over recent years both in radial velcocity surveys \citep[e.g.,][]{sanmoutsa16} and in the {\it Kepler} dataset \citep[e.g.,][]{Dong13,petmarwin18}. In particular, gas giants rise in number out to at least $\sim$300 days, unlike sub-Neptunes whose population plateaus beyond $\sim$10 days. \citet{leechi17} demonstrated that such a flat occurrence rate profile is most likely produced by the final assembly of cores that occurred largely in-situ. In situ oligarchic assembly and subsequent orbital stability demands planets to be spaced apart by some multiple of their mutual Hill radii \citep[e.g.,][]{kokida12} which naturally produces a uniform distribution of log orbital periods.\footnote{ The fall-off of the number of sub-Neptunes inside $\sim$10 days is explained by associating disk edges with the co-rotation radius with young stars \citep[see also][]{mulpasapa15}.} More importantly, small planets show no pile-up near the inner edges of their protoplanetary disks, unlike their larger counterparts. This feature is difficult to explain in the large-scale migration paradigm. What gives rise to the different shape of the orbital period distribution between the big gas giants and small sub-Neptunes? In a standard solar nebula, the migration timescale is longer farther out from the star so does the overall rise in the occurrence rate of gas giants point to their migratory history?

We answer this question by revisiting disk-induced migration of cores undergoing envelope accretion, taking into account planetary gap-opening in viscous disks. We restrict our attention to viscous-$\alpha$ disks ($\alpha\geq 10^{-3}$) because laminar environments ($\alpha< 10^{-4}$) preclude large-scale migration. To better intuit the interplay between the three processes in setting the giant planet occurrence rate, we distill our model into bare-bone essentials. The occurrence rate's sensitivity to atmospheric dust content, as well as the disk and planet initial conditions, is also probed. Our model ingredients are described in Section \ref{sec:methods}, and we present our results in Section \ref{sec:results}. After placing the results in broader context with the observations, we close in Section \ref{sec:discussion} by outlining their implications.

\subsection{The Analytic Occurrence Rate Profile}
\label{subsec:analy_orp}
We begin by deriving an analytic model of the giant planet occurrence rate. Our aim in this section is to sketch how the occurrence rate profile scales with orbital period under the assumption that planets are transported radially via disk-induced migration. Through this analytic derivation, we illustrate, in isolation, the effect of migration on the final orbital period distribution. Readers interested in the full numerical calculation---on which all results of this paper is based---should skip to Section \ref{sec:methods}. In this toy model we compute the evolution of an ensemble of equal-mass cores subject to disk-induced migration, in the absence of mass growth by gas accretion or gap opening by planet-disk interaction.

With the distribution of initial orbital period $dN/d\log P_i$ as an input parameter, we can write the planet occurrence rate as a function of final orbital period $P$ as

\begin{equation}\label{equation:occ_rate}
\frac{dN}{d\log P}=\frac{dN}{d\log P_{i}}\frac{P}{P_{i}}\frac{d P_{i}}{d P}.
\end{equation}

We relate $P_i$ and $P$ by tracking the evolution of orbital period under disk-planet tidal interaction. The disk gas surface density $\Sigma$ and temperature $T$ are assumed to follow radial power-laws: $T\propto a^{-\gamma}$, with $a$ the orbital radius and $\Sigma\propto a^{-\beta}\exp({-t / t_{\rm disk}})$, with $t_{\rm disk}$ the disk dissipation timescale. The initial period distribution is assumed logarithmically flat ($dN/d\log P_{i}$ $\propto P_{i}^{0}$), reflecting cores separated by a multiple number of their mutual Hill radii through oligarchic assembly \citep[e.g.][]{kokida12}.

Cores are transported radially according to

\begin{equation}\label{equation:adot1}
    \dot{a}=\frac{2\Gamma_{\rm net}}{M_{\rm p}\Omega a},
\end{equation}
with $\Omega$ the Keplerian orbital frequency at $a$, $M_{\rm p}$ the planet mass (assumed constant here), and $\Gamma_{\rm net}$ the net tidal torque via disk-planet interaction (e.g., \cite{klenel12,kantanszu18}):

\begin{equation}
    \Gamma_{\rm net}=-f\frac{G^2 M_{\rm p}^{2} \rho}{c_{s}^{2}}H,
\end{equation}
where $G$ is the gravitational constant, $H=c_{s}/\Omega$ is the disk scale height, $c_{s}$ is the disk sound speed, $\rho=\Sigma / H$ is the disk volumetric density, and $f$ is a numerical prefactor. This $f$ is typically constant throughout the disk, depending linearly on $\beta$ and $\gamma$ \citep[e.g.][see our equations \ref{equation:lindlad}, \ref{equation:corotation} for our treatment]{paabarcri10,danlub10}. In this section, we seek only the scaling between the occurrence profile and period, so the normalizations of $f$ and other prefactors are not relevant. Reformulating equation \ref{equation:adot1} in terms of $a$, we arrive at,

\begin{equation}\label{equation:adot_barecores}
    \dot{a}=-\xi (a/a_{0})^{1/2-\beta+\gamma}\exp({-t / t_{\rm disk}}),
\end{equation}
with $a_{0}$ the radius at which our power laws are normalized, the constant prefactor $\xi=2f a_{0}\Omega_{0}(M_{\rm p} / M_{\star})(\Sigma_{0}a_{0}^{2} / M_{\star})(a_{0}\Omega_{0} / c_{s_{0}})^{2}$, $M_{\star}$ the host star mass, and $\Omega_{0},\Sigma_{0}$ and $c_{s_{0}}$ the orbital frequency, gas surface density, and sound speed at $a_{0}$.

Re-writing equation \ref{equation:adot_barecores} in terms of periods  and plugging it into equation \ref{equation:occ_rate}, we arrive at

\begin{equation}\label{equation:occ_slope}
    \frac{d N}{d \log P} \propto P^{\frac{2}{3}(1/2+\beta-\gamma)}.
\end{equation}
In steady state, the number flux of planets is conserved across each period bin, and $dN/d\log P$ simply scales with the migration time. We note that the index, $1/2+\beta-\gamma$, is wholly dependent on disk structures and in reality will be affected by e.g., substructures and perturbations.

\begin{figure}
\epsscale{1.2}
\plotone{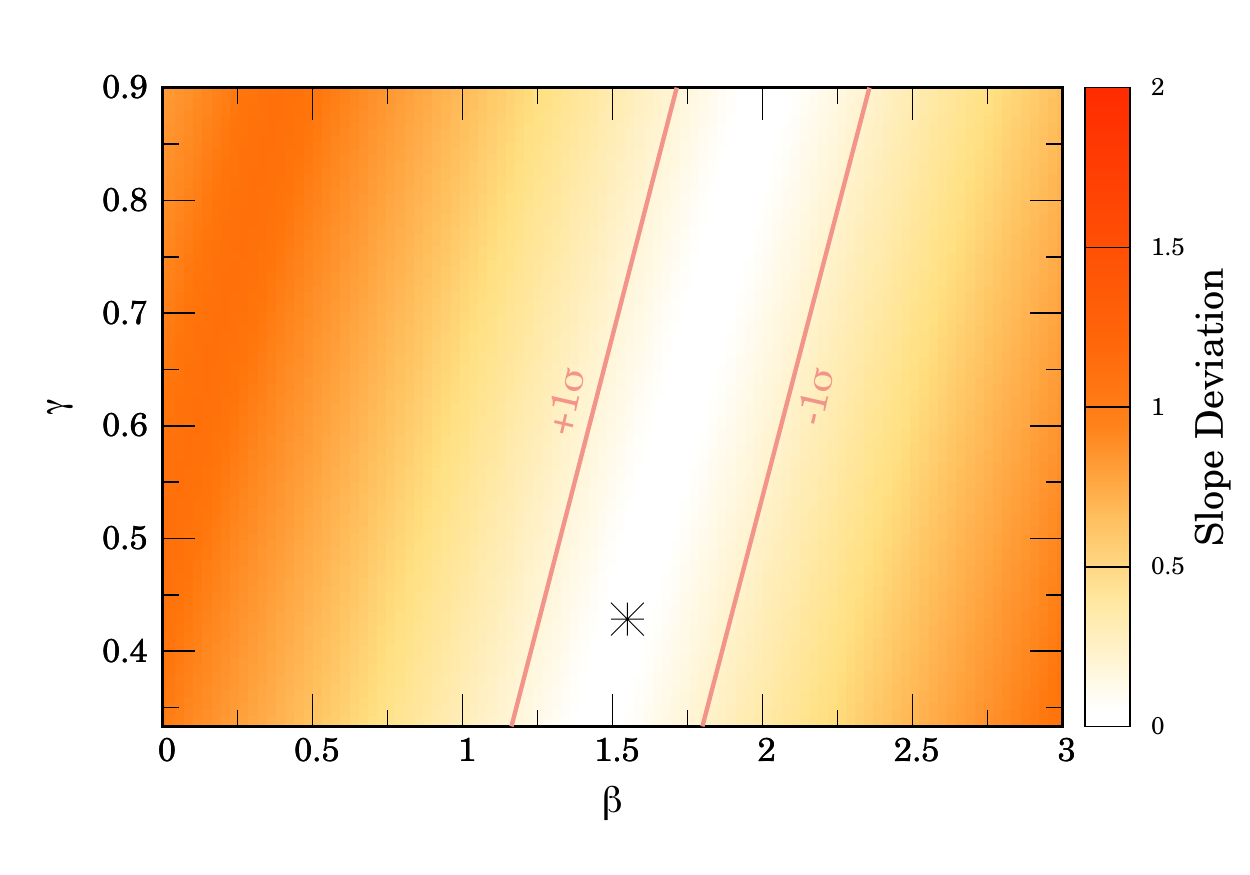}
\caption{The deviation of the analytic occurrence rate (equation \ref{equation:occ_slope}) from the slope of the occurrence rate of Jupiters from \cite{petmarwin18}, as a function of disk surface density and temperature profiles ($\Sigma\propto a^{-\beta},T\propto a^{-\gamma}$). The $\pm 1 \sigma$ upper and lower frontiers indicate where the slope agrees with the data within the errorbars from $\sim$7 to $\sim$300 days. The star locates the approximate ``best-fit'' value for $\beta$ given $\gamma=3/7$, which we adopt in our fiducial disk model. The ``best-fit'' $\beta=1.55$, and is very close to that of the Minimum-Mass Extrasolar Nebula \citep[$a^{-1.6}$;][]{chilau13}, and the Minimum-Mass Solar Nebula \citep[$a^{-1.5}$;][]{hay81}. The deviation is the absolute value difference between the slope in equation \ref{equation:occ_slope} and the Jupiter occurrence data from \cite{petmarwin18}, excluding data points without errors and those at $P<$7 days. \label{figure:bg_heatmap}}
\end{figure}

In Figure \ref{figure:bg_heatmap}, we display the deviation of our analytic result from the slope of the observations for various disk power law indices. The deviation is the absolute value difference between the slope given by equation \ref{equation:occ_slope} and that of the occurrence data from \cite{petmarwin18}. For this calculation we restrict our attention to the data points at periods $\geq$7 days, and exclude the data points without error bars at $\sim$13 and $\sim$75 days. The model slope is consistent with observations within the error bars for a narrow range of $\beta$, spanning approximately 1.25--2.25. For $\gamma = 3/7$ (consistent with a passive disk; \citealt{chigol97}), the best-fit $\beta=1.55$ which we take to describe our fiducial disk. This lies very close to that of the Minimum-Mass Extrasolar Nebula \citep[$a^{-1.6}$;][]{chilau13}, and the Minimum-Mass Solar Nebula \citep[$a^{-1.5}$;][]{hay81}. \footnote{We note that accounting for disk winds and magnetohydrodynamic processes in disks generally predict shallower slope, but these studies all require some assumed initial density profile which range from $a^{-0.5}$ to $a^{-2.2}$ \citep[e.g.,][]{Bai11,baisto13,greturnel15}. Over time, the inner disk shallows and sometimes carve out an inner hole depending on the assumed strength of disk winds \citep[e.g.,][]{Suzuki16} but there remains uncertainties in the appropriate initial conditions and disk microphysics \citep[e.g.,][]{bai17}. We focus our attention to the gas giants inside 1 AU but these inner regions of protoplanetary disks cannot yet be characterized with e.g., radio observations.}

\subsection{Order of Magnitude Estimates of Relevant Timescales}\label{subsec:timescales}

The simple analytic model derived above assumes that cores complete their migration before accreting enough gas to emerge as gas giants. Complicating this picture is the possibility that migrating cores may simultaneously accrete gaseous envelopes, which may in turn lead to gaps being carved out of the surrounding disk by the growing planet. Here we show that the timescale for gas accretion is indeed commensurate with the migration timescale for regions in the disk our model planets are typically born in. This means that in order to test whether large-scale migration is responsible for the giant planet occurrence rate, we first need to understand how migration and accretion evolve in tandem. Planets open gaps over much shorter timescales, and so the effect of gaps on migration also needs to be taken into account. 

An order of magnitude estimate of the migration timescale yields,

\begin{align}\label{equation:mig_time}
    t_{\rm mig} & \sim a/\dot{a} \nonumber \\
    & \sim 1.5\cdot 10^{4}\,{\rm yr}\,\bigg(\frac{a}{\rm AU}\bigg)^{1.6}\bigg(\frac{20 M_{\oplus}}{M_{\rm p}}\bigg)\bigg(\frac{M_{\star}}{M_{\odot}}\bigg)^{2}
\end{align}
in which we've taken $a_{0}=0.1$ AU, $\beta=1.55$, $\gamma=3/7$, $T_{0}=10^{3} \rm ~K$, $c_{s_{0}}=1.8\cdot 10^{5}$ cm  s$^{-1}$, $f=1$, $t=0$, and $\Sigma_{0}=4\cdot 10^{4}$ g cm$^{-2}$, one order of magnitude lower than the Minimum-Mass Extrasolar Nebula computed by \citet{chilau13}.

Planets with masses above the thermal mass (i.e., their Hill radii, $R_{\rm Hill}=a(M_{\rm p}/3 M_{\star})^{1/3}$, exceed the local disk scale height $H$) perturb the surrounding disk and carve out deep gaps. The time required for a planet to induce order-unity perturbations in the disk and carve out a gap is given by \citep{raf02,funlee18}

\begin{align}
    t_{\rm gap} & \sim \frac{t_{\rm cross}}{\lambda_{\rm t}} \nonumber \\
    & \sim 6\bigg(\frac{M_{\rm p}}{M_{\rm th}}\bigg)^{-14/5}\bigg(\frac{H}{a}\bigg)^{-1}\Omega^{-1} \nonumber \\
    & \sim 4\cdot 10^{2}\,{\rm yr}\,\bigg(\frac{a}{\rm AU}\bigg)^{3.6}\bigg(\frac{20 M_{\oplus}}{M_{\rm p}}\bigg)^{14/5}\bigg(\frac{M_{\star}}{M_{\odot}}\bigg)^{-7/5},
\end{align}
where $t_{\rm cross}$ is the time for disk gas to drift across the gap in the rest frame of the planet, and the $\lambda_{\rm t}$ parameter describes the effect of planetary tidal torques on the disk surface density distribution \citep{raf02}. We take $M_{\rm th}\approx 56 M_{\oplus}(a/\rm AU)^{6/7}$ as the thermal mass.

Gas giants are the outcomes of runaway gas accretion.
Using the accretion rates derived by \cite{leechiorm14}, corrected for the reduction in accretion rate by advection \citep{lamle17,funzhuchi19}, the runaway time reads

\begin{align}
    t_{\rm acc} &\sim 1.5\,{\rm Myr}\,\bigg(\frac{10^{3}\rm K}{T}\bigg)^{-3.75}\bigg(\frac{M_{\rm core}}{20 M_{\oplus}}\bigg)^{-2.5} \nonumber \\
    & \sim 4\cdot10^{4}\,{\rm yr}\,\bigg(\frac{\rm AU}{a}\bigg)^{1.6}\bigg(\frac{20 M_{\oplus}}{M_{\rm core}}\bigg)^{2.5}.
\end{align}

Our choice to normalize the timescales at 1 AU is motivated a posteriori by our numerical calculations which find warm and hot Jupiters to be generated from $\sim$1-2 AU for regions of our parameter space which best recover the data. The migration and accretion timescales are within an order of magnitude at these distances for planets of a few tens of Earth masses, meaning their migration and accretion occur simultaneously, in the presence of perturbations to the disk due to planetary torques. This requires a numerical treatment of giant planet migration theory tracking gas mass growth and disk-planet interaction in tandem.

\section{Methods} \label{sec:methods}
We use a Monte Carlo (MC) approach to build model occurrence rates of Jupiters. For each combination of model parameters, we draw 10$^5$ systems, each harbouring one planet whose evolution in mass and orbital period we follow. Each of the 10$^{5}$ disk-planet systems in a model contain the same initial core mass. We vary the core mass and disk gas surface density (through use of a depletion factor $d$) between models, defining a parameter space spanning $M_{\rm core}=$ 10--30 $M_{\oplus}$ and $d=$ 0.001--1. From the base model outlined in Section \ref{subsec:analy_orp}, we progressively add different physics describing interactions between the planet, the disk, and the host star, adding complexity to our planet population. We identify the underlying physics and related parameter values that successfully reproduce the observed shape of the occurrence rate of Jupiter-sized planets ($\geq 100 M_{\oplus}$). We then ask whether these are likely to be physically realized. 
The following sections outline the various physics reflected in our models. In Section \ref{subsec:disks} we lay out our choice of disk model, \ref{subsec:migration} and \ref{subsec:augap} describe how we treat planetary migration, \ref{subsec:accretion} discusses our model of envelope accretion, and \ref{subsec:tides} summarizes our treatment of planet-star tidal interaction.

\subsection{Disk Model $\&$ Magnetospheric Truncation} \label{subsec:disks}
Guided by our simple analytic solution from Section \ref{subsec:analy_orp}, we choose $\beta$=1.55, and $\gamma$=3/7 as our fiducial disk profile. We choose for the disk evaporation time $t_{\rm visc}$=0.7 Myr \citep{oweerccla11}. 

Inside 1 AU, the temporal and spatial evolution of the underlying disk is described as follows:
\begin{equation}\label{equation:passive_disk}
    \frac{\Sigma_{\rm bg}(a,t)}{\textrm{g cm$^{-2}$}}=d\cdot\left\{\begin{array}{ll}
    3\cdot10^{4}(\frac{a}{0.2 \mathrm{AU}})^{-\beta}\Big(\frac{t}{t_{\rm visc}}+ 1\Big)^{-1.54}, & t<t_{\rm visc} \\
    C\cdot(\frac{a}{0.2 \mathrm{AU}})^{-\beta}\exp(-\frac{t}{\rm t_{visc}}), & t>t_{\rm visc}, \\
    \end{array} \right.
\end{equation}
with $a$ the semi-major axis, $d \leq 1$ a density depletion factor yielding gas-poor disks (with the normalization at $d$=1 corresponding to typically observed accretion rates of T Tauri stars), and $C \sim 3 \cdot 10^4$ a normalization prefactor for continuity between branches. The first branch ($t<t_{\rm visc}$) corresponds to a steady-state, viscously evolving disk \citep{lynpri74,harcalgul98}, while the second branch ($t>t_{\rm visc}$) reflects a disk evolving after a photoevaporative wind carves out a gap at $\sim$1 AU and decouples the inner disk from the outer disk \citep{oweerccla11}. The temperature profile is normalized as $T_{\rm disk} = 10^{3} ~ \textrm{K}(a/0.1 \textrm{AU})^{-3/7}$.

Following the work of \citet{leechi17}, we assume the disks are magnetospherically truncated at co-rotation with respect to the star so that the innermost disk edge $P_{\rm in}=P_\star$ where $P_\star$ is the stellar rotation period  \citep[e.g.,][]{gholam79,koeari91,ostshu95,lonromlov05,romowo16}. We sample $P_\star$ from the observed distribution of pre-main sequence stellar rotation periods from Orion Nebula Cluster \citep[ONC, $\sim$1 Myr;][]{herbaimun02,rodmuneis09}, limiting the sample to stars with masses 0.5--1.4$M_\odot$, representative of FGK dwarfs. There remains uncertainty on how long the disk can remain locked with the star. For example, as the disk loses its mass and the magnetosphere extends beyond the co-rotation radius, the disk gains angular momentum from the star and pushes out (the so-called ``propeller'' regime; \citealt{romowo16}). We find observational support for prolonged disk locking in dipper stars, stars of a few Myrs old exhibiting quasi-periodic attenuation due to the presence of dust clumps and/or accreting disk material lifted out of the disk midplane at corotation \citep{stacodmcg15,ansgairap16,bodquians17}. Furthermore, gas and larger solids that are invisible to our telescopes may extend to the corotation radius even if smaller dust grains truncate outside \citep[e.g.,][]{Eisner05}.

\subsection{Planetary Migration $\&$ Gap-Opening} \label{subsec:migration}
Planets are transported radially according to equation \ref{equation:adot1}, with $\Gamma_{\rm net}$ the combined Lindblad and co-rotation torque exerted on the planet \citep[e.g.,][]{klenel12,kantanszu18}. Assuming locally isothermal disks, we adopt the Lindblad ($\Gamma_{\rm L}$) and co-rotation ($\Gamma_{\rm C}$) components from \citet[][see their equations 14, 17, 45, and 49]{paabarcri10}:
\begin{equation}\label{equation:lindlad}
    \frac{\Gamma_{\rm L}}{\Gamma_{0}(a)}=-(2.5-0.1\beta+1.7\gamma)b^{0.71},
\end{equation}
\begin{equation}\label{equation:corotation}
    \frac{\Gamma_{\rm C}}{\Gamma_{0}(a)}=1.1(1.5-\beta)b + 2.2\gamma b^{0.71} -1.4\gamma b^{1.26},
\end{equation}
with $b=2/3$ a gravitational softening parameter, and $\Gamma_{0}$ the characteristic torque on a planet,
\begin{equation}\label{equation:characteristic_torque}
    \Gamma_{0}=2\pi\int_{a_{\rm in}}^{a_{\rm out}}\Lambda \Sigma_{\rm bg}a^{\prime}da^{\prime},
\end{equation}
where $a_{\rm in}$ and $a_{\rm out}$ are the disk inner and outer edges, and $\Lambda$ is the characteristic torque per unit mass \citep{linpap86}:
\begin{equation}\label{equation:lambda_torque}
    \Lambda=f_{\Lambda}\bigg(\frac{M_{\rm p}}{M_{\star}}\bigg)^{2}\frac{G M_{\star}}{2a^{\prime}}\bigg(\frac{a^{\prime}}{\Delta_{\rm p}}\bigg)^{4}F(a^{\prime}),
\end{equation}
with $f_{\Lambda}$ a dimensionless prefactor, $\Delta_{\rm p}=\max(H(a),|a^{\prime}-a|)$, $H(a)$ the disk scale height evaluated at $a$, and $F(a^{\prime})=(a^{\prime}-a)/H(a)$ for $|a^{\prime}-a|\leq H(a)$, and $F(a^{\prime})=\rm sign(a^{\prime}-a)$ otherwise. We take $a_{\rm out}=100$ AU, and fix $f_{\Lambda}$=0.078 in order to match the empirical migration rate estimates of \cite{kantanszu18} (this choice of $f_{\Lambda}$ yields agreement with their equation 16). There are a number of caveats associated with our adoption of equations \ref{equation:lindlad}, \ref{equation:corotation}. We assume a locally isothermal equation of state as treated by \cite{kantanszu18}, and we neglect thermal diffusion \citep{paabarkle11} and distance-dependent disk opacities \citep{masmorcri06,colnel16}. A thorough review of our model's caveats and their potential impact on our results is presented in Section \ref{sec:caveats}.

Planets with super-thermal masses perturb the surrounding disk and carve out deep gaps. In steady-state, the one-sided Lindblad torque pushing gas outward is balanced by the viscous torque pulling gas inward, giving an estimate of the density depletion factor at the bottom of the gap \citep{funshichi14}:
\begin{equation}\label{equation:depletion_k}
    K=\bigg(\frac{H}{a}\bigg)^{-5}\bigg(\frac{M_{\rm p}}{M_{\star}}\bigg)^{2}\alpha^{-1},
\end{equation}
with $\alpha$ the Shakura-Sunyaev viscous parameter. We consider $\alpha=10^{-3}$, self-consistent with an accretion rate of $4\cdot 10^{-8}M_{\odot}$  yr$^{-1}$ \citep{oweerccla11}. The empirically determined gap depth derived by \cite{dufmac13} then reads
\begin{equation}\label{equation:gap_depth}
    \Sigma_{\rm neb}=\frac{\Sigma_{\rm bg}}{1+0.034K},
\end{equation}
with $\Sigma_{\rm neb}$ the gas density at the bottom of the gap. We use the steady-state migration torque encompassing the no-gap and deep-gap regimes as empirically derived by \cite{kantanszu18}:
\begin{equation}\label{equation:net_torque}
    \Gamma_{\rm net}=\frac{\Gamma_{\rm L}+\Gamma_{\rm C}\exp(-K/K_{\rm t})}{1+0.034K},
\end{equation}
where the exponential term indicates the quenching of co-rotation torque in the deep-gap limit. \footnote{\cite{chezhali20} report order-unity corrections to the gas density at the bottom of the gap computed by \cite{kantanszu18} for some disk configurations.} We follow \cite{kantanszu18} in setting $K_{\rm t}$=20 to be the gap depth at which the corotation torque becomes ineffective. In the no-gap limit, i.e. when $K\ll K_{\rm t}$ and $0.034K\ll 1$, the net torque is the sum of the Lindblad and co-rotation components (equations \ref{equation:lindlad} and \ref{equation:corotation}). In the deep-gap limit, i.e. when $K\gg K_{\rm t}$ and $0.034K\gg 1$, the torque reduces to $\Gamma_{\rm L}/(0.034K)$. This treatment of the corotation torque saturation applies when $10^{-4}<\alpha< 10^{-2}$; below $10^{-4}$ the disk behaves as if it were inviscid \citep[e.g.][]{funchi17,ginsar18,funlee18}, and above $10^{-2}$, $K_{\rm t}\sim\infty$ for $10<K<10^{3}$ \citep{kantanszu18}.

When a planet migrates to the inner disk edge, we set $\dot{a}=0$, and set its final location to $P_{\rm final} = 0.5P_{\rm in}$. In other words, we assume planets which reach the inner edge always end at the inner 2:1 resonance. Whether planets can reach the inner 2:1 resonance depends on the shape of the inner edge, the planet mass and eccentricity, whether eccentricities can be excited by companions, as well as the migration prescription \citep[e.g.,][]{bramatmut18}. In the vicinity of the inner edge, an asymmetric torque develops due to the one-sided surface density contrast. This torque becomes very negative (akin to the torque at the $\sim$1 AU gap in Figure \ref{figure:asym_torque}, but differing in sign), shortening significantly the migration timescale. This supports our assumption of instantaneous placement of planets at the inner 2:1 resonance upon their arrival at the disk inner edge.

More generally, planets which reach the innermost disk edge may park interior to the edge \citep{goltre80,linbodric96}, at the disk edge \citep{tantakwar02,masdankle06,terpap07}, or exterior to the edge \citep{tsa11,Miranda18}. Adjusting the choice of the final location does not change the slope of our model occurrence rates outside $\sim$10 days, nor does it affect our conclusions regarding how planetary gap opening and accretion set the occurrence rate. The observed peak of the hot Jupiter occurrence rate is at $\sim$4--5 days, close to 2:1 resonance with the peak of the ONC $P_{\rm in}=P_\star$ distribution. Our choice of $P_{\rm final}$ is the most favorable condition for migration theory to produce hot Jupiters. We present our model setup as a strong test of migration theory as even with this optimistic configuration, we are unable to find a region of parameter space that can consistently explain the observed orbital period and mass distribution of hot and warm Jupiters.

We draw pre-migration planet periods randomly and uniformly in log period from $\rm P_{\rm min}$ to $\rm P_{\rm out}=365\cdot 10^{3}$ days (100 AU). A uniform logarithmic distribution reflects the spacing of planetary systems whose cores undergo oligarchic assembly. These planets maintain separations of a multiple number of their mutual Hill radii in the late stages of core formation \citep[e.g.][]{kokida12}. For each combination of disk parameters and core mass, we choose P$_{\rm min}$ so that the smallest final period in the model ensemble of planets is $\sim$1 day. Setting P$_{\rm min}$ this way ensures that we sample the hot Jupiters without overproducing them; sampling smaller periods often produces a pile-up of hot Jupiters at the inner disk edge, inconsistent with the modest increase in occurrence at $\sim$4 days. This gives the large-scale migration hypothesis the most favourable conditions to succeed, strengthening our test of the theory. We then integrate equation \ref{equation:adot1} over 10 Myr. For most of our model parameters, $P_{\rm min} \gtrsim 365$ days. Models with high core masses in depleted disks (i.e., $d\lesssim 10^{-1}$) require $P_{\rm min} < 365$ days.

\subsection{A Photoevaporative Gap in the Disk}\label{subsec:augap}
Photoevaporative mass-loss in disks, coupled with viscous evolution, carves out a gap in the disk at $\sim$1 AU \citep{clagensot01,oweerccla11}. The asymmetry in disk gas surface density in the vicinity of the gap gives rise to an asymmetric tidal torque; planets just interior to the gap feel an increased outward torque due to the preponderance of disk material interior to the planet than exterior. We compute migration due to the asymmetric torque after the gap has been cleared at $t_{\rm AU \rm gap}=3$ Myr until the end of our simulations. We stop the migration for planets that are left beyond 1 AU as they are outside the range of orbital periods of our interest. Hereafter we refer to the photoevaporative gap at $\sim$1 AU as `the 1 AU gap', and refer to gaps carved out by planets at their locations as `planetary gaps'.

\begin{figure}[]
\epsscale{1.2}
\plotone{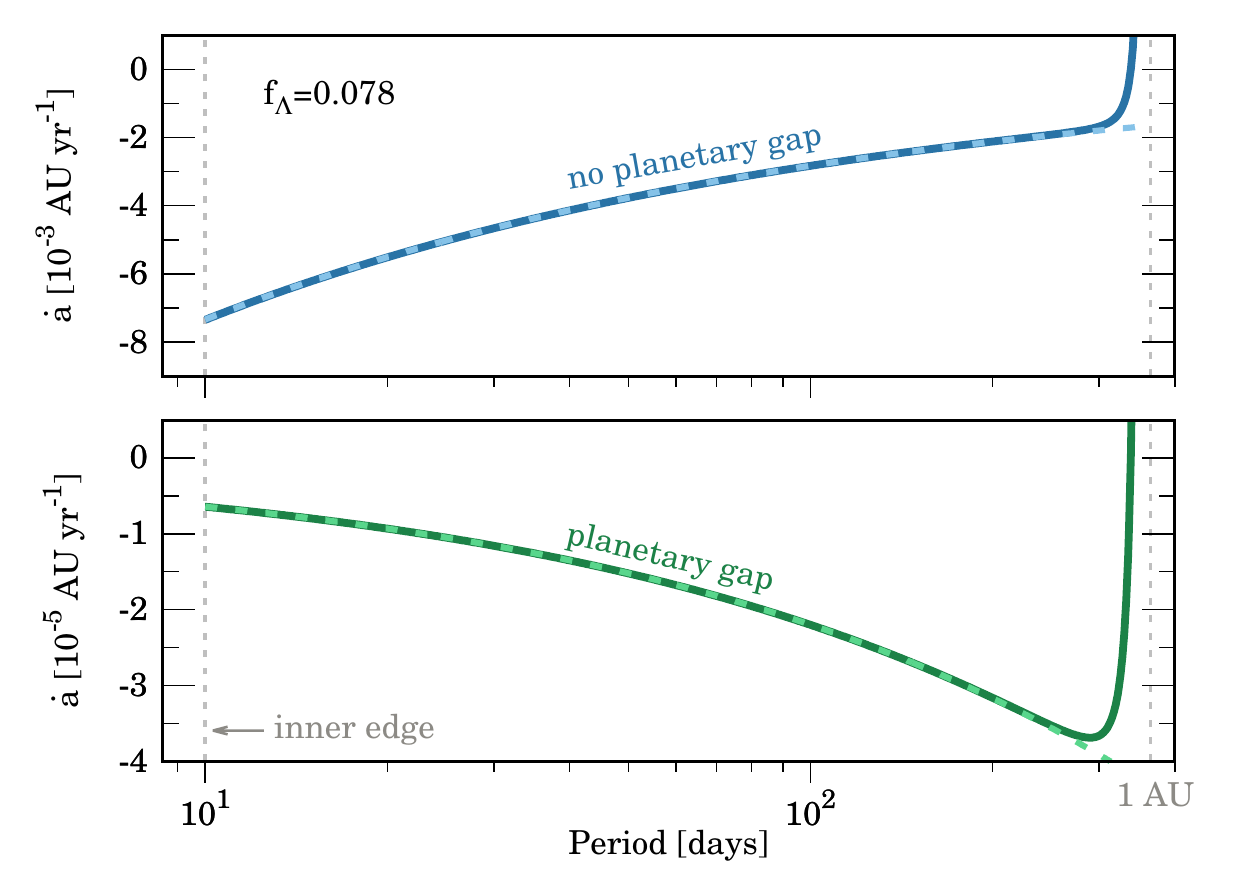}
\caption{Sample migration rates for a 100 $M_{\oplus}$ planet without including planetary gap-opening (top panel) and including gap-opening (bottom panel), accounting for the asymmetric torque near the outer disk edge (solid curves) and not (dashed curves). The effect of the photoevaporative gap at 1 AU is to introduce large positive torques at the disk edge, due to the asymmetric disk surface density there. These push planets outwards at the outer edge of the disk. Planets are halted once they reach the disk inner edge, then placed at the inner 2:1 resonance. The inner edge is taken to be at 10 days here as an example. For illustration, we set $d=1$ and $t=0$ (gas-full disk at time zero; see equation \ref{equation:passive_disk} for their definitions). The constant $f_{\Lambda}$ (see equation \ref{equation:lambda_torque}) is tuned to match the empirical migration rate estimates of \cite{kantanszu18}. Without opening gaps, planets accelerate as they migrate inwards. Once they open gaps, they preferentially slow down as they move closer to the central star.  \label{figure:asym_torque}}
\end{figure}

Planets evolve according to equation \ref{equation:adot1}, with the net torque given by equations \ref{equation:net_torque}, \ref{equation:lindlad}, and \ref{equation:corotation} as usual. Before simulation times reach $t_{\rm AU \rm gap}$, we use equation \ref{equation:characteristic_torque} with $a_{\rm out}$=100 AU for the characteristic torque. Once simulation times reach $t_{\rm AU \rm gap}$, we set $a_{\rm out}$= 1 AU to account for the asymmetric torque. The pre- and post- $t_{\rm AU \rm gap}$ migration rates agree in middle regions of the disk, far from the disk edges (see Figure \ref{figure:asym_torque}).

In transitioning from the no-gap to the deep-gap limit, migration rates shift from being inversely proportional to the disk aspect ratio ($\propto (H/a)^{-2}a^{3}\Omega a^{-\beta}\propto a^{-0.6}$) to super-linearly proportional ($\propto (H/a)^{3}a^{3}\Omega a^{-\beta}\propto a^{0.8}$), due to the gap depletion factor's steep inverse dependence on $H/a$ (only disk temperature profiles steeper than $T_{\rm disk} \propto a^{-0.97}$ produce migration rates inversely proportional to $a$ in the deep-gap limit for $\beta=1.55$). 

These dependencies imply that gap-opening planets are expected to pile up at close-in distances. In order to reconcile migration theory with observations---which show fewer gas giants closer to the star---planet masses cannot be uniform across all orbital periods.

\begin{deluxetable*}{cCCCCC}
\tablecaption{Parameter Values Used in Our Models. \label{tab:parameters}}
\tablecolumns{4}
\tablenum{1}
\tablewidth{0pt}
\tablehead{
\colhead{Parameter} &
\colhead{Definition} &
\colhead{Units} &
\colhead{Values} & 
\colhead{Reference}}
\startdata
$\beta$ & $\Sigma$\propto a$^{-\beta}$ & \nodata & 1.55 & \nodata \\
$\gamma$ & T$_{\rm disk}$\propto a$^{-\gamma}$ & \nodata & 3/7 & (1) \\
$M_{\rm core}$ & \rm core ~ \rm mass & $M_{\oplus}$ & 10-30 & \nodata \\
$d$ & $\Sigma$ ~ \rm depletion ~ \rm factor & \nodata & 10$^{-3}$-1 & \nodata \\
P$_{\star}$ & \rm disk ~ \rm inner ~ \rm edge = P$_{\rm in}$ = \rm P$ _{\star}$ & \nodata & \rm Orion ~ \rm Nebula ~ \rm Cluster ~ (ONC) & (2), (3) \\
$t_{\rm AU \rm gap}$ & \rm photoevaporative ~ \rm gap ~ \rm time & \rm Myr & 3 & (4) \\
$\alpha$ & \rm Shakura-Sunyaev ~ \rm viscous ~ \rm parameter & \nodata & 10$^{-3}$ & (4) \\
$\sigma_{\star}$ & \rm stellar ~ \rm tidal ~ \rm dissipation ~ \rm constant & \rm g^{-1}\rm cm^{-2}\rm s^{-1} & 5\cdot 10^{-66} & (5) \\
\enddata
\tablecomments{$\beta$ is 
the best-fit parameter for our our analytic occurrence rate profile from Section \ref{subsec:analy_orp}, assuming $\gamma=3/7$. 
$\Sigma$ is normalized such that $d=1$ corresponds to typically observed accretion rates of T Tauri stars $\sim 10^{-8} M_{\odot} $yr$^{-1}$
for $\alpha=10^{-3}$.
}
\tablerefs{(1) \cite{chigol97}, (2) \cite{herbaimun02}, (3) \cite{rodmuneis09}, (4) \cite{oweerccla11}, (5) \cite{han10}}
\end{deluxetable*}

\subsection{Envelope Accretion}\label{subsec:accretion}
In tandem with radial migration, gas accretion onto these cores is followed using the approach of \citet{lee19}. We account for three modes of accretion---accretion by cooling, local hydrodynamic flows, and the global, disk gas accretion rate. We describe our treatment of each below.

\subsubsection{Cooling-Limited Accretion: Dusty and Dust-free}\label{subsubsec:cooling}
In this regime, the rate of accretion is set by the rate of cooling of the accreted gas \citep{leechi15}. The rate of gas cooling in turn is determined by the thermodynamic properties of the radiative-convective boundary; in particular, it is sensitive to the opacity \citep{ikonakemo00} and whether it is dominated by dust grains or gaseous molecules \citep{leechiorm14,pisyoumur15}. We explore the two maximal limits: dusty envelopes with interstellar medium grain size distribution, and dust-free envelopes.

For dusty gas accretion, we adopt the analytic scaling relationship between gas mass, core mass, and time derived by \cite{leechi15}, corrected for the weak dependence on nebula gas density \citep{leechi16}, and the factor of $\sim3$ shrinkage of the bound radius by hydrodynamic advection \citep{lamle17,funzhuchi19}:
\begin{multline}\label{equation:dusty_gcr}
    \frac{M_{\rm gas}}{M_{\rm core}}=0.09\bigg(\frac{\Sigma_{\rm neb}}{13 \rm g \rm cm^{-2}}\bigg)^{0.12}\bigg(\frac{M_{\rm core}}{20 M_{\oplus}}\bigg)^{1.7}\bigg(\frac{t}{0.7 \rm Myr}\bigg)^{0.4}\\
    \times\exp\bigg[\frac{t}{2.2 \rm Myr}\bigg(\frac{M_{\rm core}}{20 M_{\oplus}}\bigg)^{4.2}\bigg],
\end{multline}
where $M_{\rm gas}$ is the gas mass, $M_{\rm core}$ the mass of the core, and $t$ the time elapsed for gas accretion. The onset of runaway accretion, when envelope self-gravity catastrophically shortens the cooling timescale, is approximated as an exponential. Differentiating equation \ref{equation:dusty_gcr} with respect to time, we have the dusty cooling accretion rate \citep{lee19}:
\begin{equation}\label{equation:dusty_mdot}
    \dot{M}_{\rm cool}=M_{\rm gas}\bigg[\frac{0.4}{t}+\frac{1}{2.2 \rm Myr}\bigg(\frac{M_{\rm core}}{20 M_{\oplus}}\bigg)^{4.2}\bigg].
\end{equation}

Similarly, for dust-free accretion,
\begin{multline}\label{equation:dustfree_gcr}
    \frac{M_{\rm gas}}{M_{\rm core}}=0.01\bigg(\frac{\Sigma_{\rm neb}}{13 \rm g \rm cm^{-2}}\bigg)^{0.12}\bigg(\frac{M_{\rm core}}{5 M_{\oplus}}\bigg)\bigg(\frac{t}{10^{4} \rm yr}\bigg)^{0.4}\\
    \times \bigg(\frac{10^{3} \rm K}{T_{\rm rcb}}\bigg)^{3/2}\exp\bigg[\frac{t}{50 \rm Myr}\bigg(\frac{10^{3} \rm K}{T_{\rm rcb}}\bigg)^{3.75}\bigg(\frac{M_{\rm core}}{5 M_{\oplus}}\bigg)^{2.5}\bigg],
\end{multline}
where $T_{\rm rcb}$ is the temperature at the radiative-convective boundary. Without the opacity provided by dust grains, the outermost atmospheric layer becomes nearly isothermal so that the temperature at the radiative-convective boundary matches that of the disk: $T_{\rm rcb}\sim T_{\rm disk}$ \citep{leechi15}. We note that these envelopes grow faster at larger orbital distances, where the disk is colder and ro-vibrational degrees of freedom of gaseous molecules freeze out, making cooling more efficient. As we will show later, this dependence on orbital distance bears critically on the mode of gas accretion that can reproduce the observed properties of gas giants.
Differentiating, we produce the dust-free cooling accretion rate:
\begin{multline}\label{equation:dustfree_mdot}
    \dot{M}_{\rm cool}=M_{\rm gas}\bigg[\frac{0.4}{t}+\frac{1}{50 \rm Myr}\bigg(\frac{10^{3} \rm K}{T_{\rm rcb}}\bigg)^{3.75}\bigg(\frac{M_{\rm core}}{5 M_{\oplus}}\bigg)^{2.5}\bigg].
\end{multline}

\subsubsection{Local Hydrodynamic Accretion}
Once the planet triggers runaway gas accretion, the cooling timescale shortens catastrophically and the accretion rate is instead limited by hydrodynamic delivery of gas. For the local flow, we take the scaling relationship reported by \cite{tanwat02}:
\begin{equation}\label{equation:hydro_mdot}
    \dot{M}_{\rm hydro}=0.29\bigg(\frac{M_{\rm p }}{M_{\star}}\bigg)^{4/3}\Sigma_{\rm neb}\bigg(\frac{a}{H}\bigg)^{2}a^{2}\Omega,
\end{equation}
with $0.29$ an empirical prefactor reported from the gas giant accretion simulations of \cite{tantan16}. This superlinear scaling with the planet mass can be understood from isothermal shocks across the envelope boundary by the inflowing gas \citep{tanohtmach12,lee19}. We note that the shock may be adiabatic in which case, $\dot{M}_{\rm hydro} \propto M_{\rm p}^{2/3}$. 

Throughout the entirety of our parameter space, we verify the radiation timescale is shorter than the orbital time (see equation 8 of \citealt{lee19}), so we assume an isothermal shock. 
In deriving equation \ref{equation:hydro_mdot}, we have assumed our planets to be super-thermal. Such an assumption is valid in the range of orbital distances of our concern. At larger distances (e.g., beyond 10s of AU), the accretion rate is governed by Bondi accretion and takes a different form \citep[see, e.g.,][]{ginchi19}. We have verified that all of our planets throughout the entire parameter space that we explore are super-thermal once they trigger runaway gas accretion.

\subsubsection{Global disk Accretion}
For global delivery of gas, we adopt the viscous spread of underlying disk gas $\dot{M}=3\pi\alpha c_{s}H\Sigma_{\rm bg}(a,t)$. For our fiducial disk indices, from equation \ref{equation:passive_disk},
\begin{equation}\label{equation:passive_global}
    \dot{M}_{\rm disk}=d\cdot\left\{\begin{array}{ll}
    4\cdot10^{-8}M_{\odot}\rm yr^{-1}\Big(\frac{t}{t_{\rm visc}}+ 1\Big)^{-1.54}, & t<t_{\rm visc} \\
    3.7\cdot10^{-8}M_{\odot}\rm yr^{-1}\exp(-\frac{t}{\rm t_{visc}}), & t>t_{\rm visc}, \\
    \end{array} \right.
\end{equation}
where $\alpha=10^{-3}$. 

For each planet, we evolve its gas mass by numerically integrating min($\dot{M}_{\rm cool}$, $\dot{M}_{\rm hydro}$, $\dot{M}_{\rm disk}$). We have verified that our method matches that of \cite{roschigin20}, who study how the gap depth evolves with gas accretion. Planetary masses and locations that obtain min$(\dot{M}_{\rm hydro},\dot{M}_{\rm disk}) = \dot{M}_{\rm disk}$ are equivalent to their consumption-limited regime, and those that obtain min$(\dot{M}_{\rm hydro},\dot{M}_{\rm disk}) = \dot{M}_{\rm hydro}$ are equivalent to their repulsion-limited regime.

We set $\dot{M}$=0 in the event that the envelope mass becomes larger than the initial disk mass reservoir, if the envelope accretion rate for any given timestep is larger than the total disk mass at that time, or if $\Sigma_{\rm neb}$ depletes by 8 orders of magnitude from its initial value, leading to prohibitively small growth by cooling \citep{leechifer18}.

\subsection{Tidal Dissipation}\label{subsec:tides}
Equilibrium tides raised by a planet on its host star cause the planet-star separation to evolve according to,
\begin{equation}\label{equation:tides}
    \dot{a}=-9\frac{M_{\rm p}M}{M_{\star}}\frac{R_{\star}^{10}}{a^{7}}\sigma_{\star}\bigg[1-\frac{\omega_{\star}}{\Omega}\bigg],
\end{equation}
with $M=M_{\rm p}+M_{\star}$ the total mass, $R_{\star}$ the stellar radius, $\sigma_{\star}$ the stellar tidal dissipation constant defined by \citet{han10}, and $\omega_\star$ the stellar rotation period. 

We set the planetary eccentricity (and hence its tidal dissipation) to zero in writing equation \ref{equation:tides}
as in gas-rich environments, dynamical friction can damp away planet eccentricities on timescales orders of magnitude shorter than the disk lifetime \citep[e.g.,][]{Papaloizou00}.
While interaction with the gas disk can in some circumstances excite eccentricities \citep[e.g.,][]{Goldreich03}, re-writing equation \ref{equation:tides} to account for tides raised on the planet and computing it for moderate eccentricities ($\sim$0.1), planetary dissipation constant $\sigma_{\rm p}=2\cdot 10^{-60}~\rm{g}^{-1}~\rm{cm}^{-2}~\rm{s}^{-1}$ \citep{han10}, and planet radius $R_{\rm p}=11 ~R_{\oplus}$ (assuming no planetary spin), tidal dissipation affects the occurrence rate profile out to $\sim$5 days, reshaping the hot Jupiter peak. This peak is pushed inwards to $\sim$3 days from $\sim$4 days, while planets inside $\sim$2 days are destroyed.

Solar mass stars evolve along the Hayashi track for the first $t_{\rm H}\sim$20-60 Myrs, so we adopt $R_{\star}=R_{\odot}(t/t_{\rm H})^{-1/3}$ and $\omega_{\star}=2\pi/P_{\star}$. We set $t_{\rm H}=20 ~\rm Myr$. After $t_{\rm H}$, we fix $R_{\star}=R_{\odot}$ and account for the stellar spin-down with $\omega_{\star} \propto t^{-1/2}$ \citep{sku72}. Following \cite{han10}, we take $\sigma_{\star}=5\cdot10^{-66} \rm g^{-1} \rm cm^{-2} \rm s^{-1}$, calibrated from systems with giant planets (see their equation 7 and Section 4.4.2). For each planet, we integrate equation \ref{equation:tides} from the end of our disk lifetime at 10 Myr to 5 Gyr. Planets which spiral inwards beyond the Roche limit, at $R_{\rm R}=3.5 R_{\star}(\rho_{\star}/\rho_{\rm J})^{1/3}$, are discarded from our model occurrence rates. The prefactor of 3.5 is supported by interpretations of the sub-Jovian desert via the destruction of planets at the Roche limit \citep[e.g.,][]{matkon16}. We take $\rho_{\star}$ to be the bulk stellar density as a function of time as the star contracts along the pre-main sequence (assuming constant stellar mass $M_{\star}=M_{\odot}$), and $\rho_{\rm J}$ the density of Jupiter.

For each model we record the final periods of our planets and construct model occurrence rates. We compare the slope of our models to the data by eye, aiming to identify broad trends and regions of parameter space that are compatible with observations. We normalize our model distributions to match the observations at 236 days. A summary of parameters and values we assign them is provided in Table \ref{tab:parameters}. 

\section{Results}\label{sec:results}

We report our results for the model occurrence rate of Jupiters below. We first outline the effect of planetary gap-opening on our model occurrence rates, showing that it flattens the slope of the occurrence rate when planets
over a range of orbital periods have roughly equal masses. Next we demonstrate how circumventing this effect to recover the correct slope necessitates a gradient in planet mass, increasing with orbital period. This can only be achieved by planets with atmospheres free of dust grains. We then extend our bare-bones model to account for varying disk and core masses, showing that under none of our initial conditions can the correct occurrence profile and mass distribution of low-eccentricity ($e<$0.1) giants be simultaneously recovered. The gradient of planet masses we find to reconcile the observed occurrence rate is too abrupt in comparison to the data - in all iterations of our model, migration predicts a bottom-heavy hot Jupiter mass distribution.

\subsection{The Effect of Planetary Gap-Opening}\label{subsubsec:occ_barecores}
\begin{figure}
\centering
\includegraphics[width=0.5\textwidth]{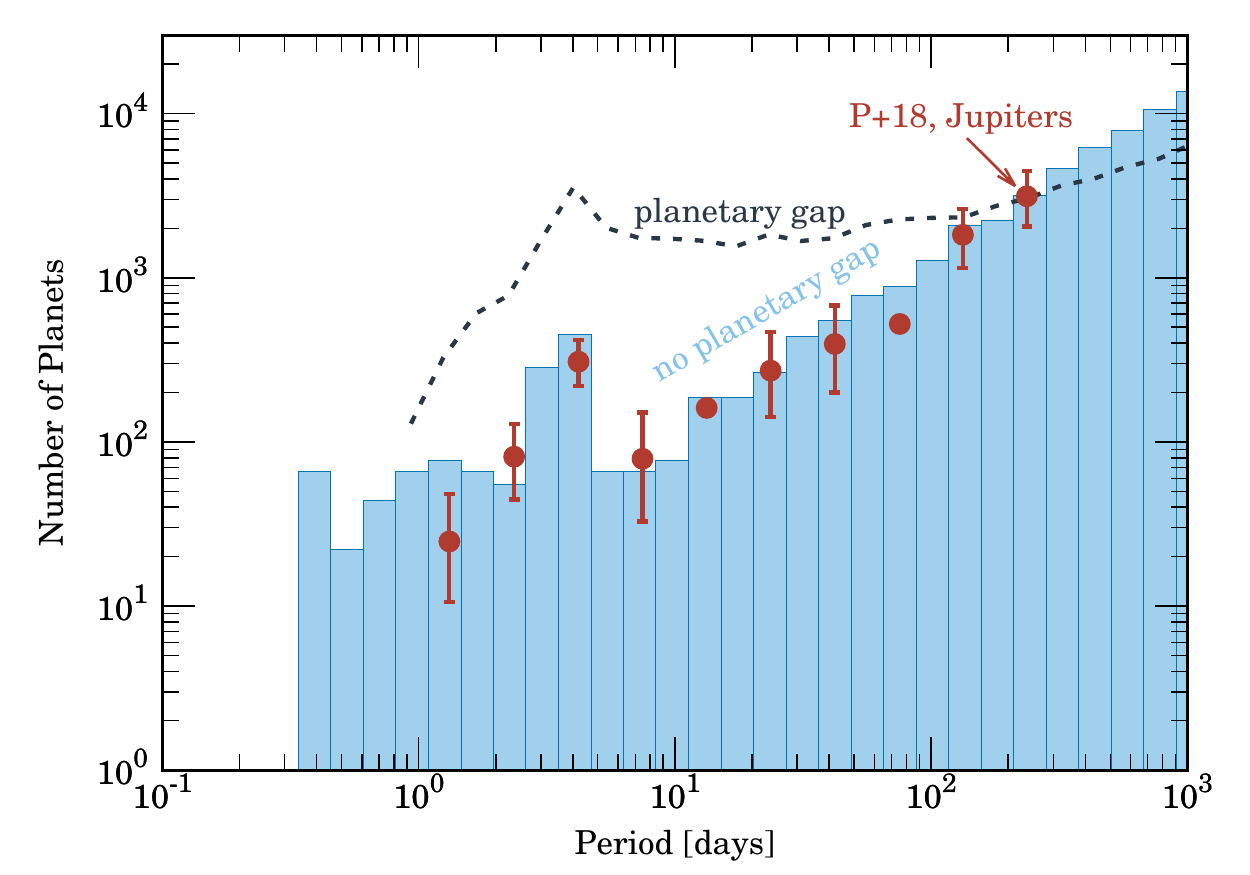}
\caption{The effect of planetary gap-opening on model occurrence rates for 18 $M_{\oplus}$ cores without gaseous envelopes using fiducial disk profiles ($\beta=1.55$, $\gamma=3/7$). The measured occurrence rate for Jupiters \citep{petmarwin18} are in red points. Cores which migrate without carving out gaps in the disk produce the orbital period distribution drawn in blue bars. Cores which carve out gaps as they move inwards through the disk are given by the navy dashed curve. Model histograms are normalized to match the data at 236 days. Disk inner edges are sampled from the rotation period distribution of the Orion Nebula Cluster \citep{herbaimun02,rodmuneis09}. Planets are set to halt their migration at the inner 2:1 resonance with these periods, giving a peak in occurrence at the observed 4 days. For a fixed mass of the planets, taking into account gap-opening by disk-planet interaction slows down the planet closer to the star and so flattens the distribution of orbital periods.
 \label{figure:barecores}}
\end{figure}

To assess the role of gap-opening in isolation, we turn off gas accretion and compare the migration of bare cores with and without planetary gap opening taken into account, in disks with fiducial profiles ($\beta=1.55$, $\gamma=3/7$).
Figure \ref{figure:barecores} demonstrates that
gap-opening flattens the occurrence rate, inconsistent with the observations.

This flattening can be easily understood from how the migration timescale scales with orbital period, since in steady state $dN/d\log P$ scales with the migration time. In Section \ref{subsec:augap}, we derived migration rates in passive disks without a planetary gap ($\dot{a} \propto a^{-0.6}$) and with a planetary gap ($\dot{a} \propto a^{0.8}$). It follows that the  time to migrate cores without opening any gap (and therefore $dN/d
\log P$) scales as $t_{\rm mig} \propto a/\dot{a} \propto a^{1.6} \propto P^1$. In the deep-gap limit, $t_{\rm mig}$ $\propto dN/d\log P$ $\propto a^{0.2} \propto P^{0.1}$, significantly shallower than the no-gap scenario. In order for the migration timescale in the deep-gap limit to scale $\propto P^{1}$, one needs $\beta\gtrsim 2$ for $\gamma\leq 1$. For our surface density normalization, these disk configurations are only marginally stable to gravitational collapse, with Toomre Q values $3\lesssim Q \lesssim 10$. In the picture that envelope accretion occurs after large-scale migration, the deep-gap occurrence rate profile fails to recover the slope of the data.

The numerical result lies between the no-gap and deep-gap solutions---our 18 $M_{\oplus}$ bare cores do not immediately carve out deep gaps, but only reach this limit once they move sufficiently far into the disk (inside $\sim$100 days, where the gap depletion factor $1/(1+0.034 K)\lesssim 0.2$). For this reason, the slope in our numerical calculation is slightly steeper than that predicted analytically.

The uptick in occurrence at 4 days is due to our sampling of the innermost disk edge from the stellar rotation periods of ONC, and our choice of resonance condition for our planets' final periods. These choices yield good overall agreement to the data for the location of the peak. Adjusting either of these conditions alters this location, so
we choose this as our ``best-fit'' setup. 

\subsection{Envelope Accretion: Dusty vs. Dust-free}\label{subsubsec:occ_accretion}
Planetary gap-opening complicates the initially simple picture suggested in our analytic solution from Section \ref{subsec:analy_orp}. When the planet masses in a model ensemble are equal
and super-thermal,
gap-opening flattens the occurrence rate profile. Our goal is to reproduce the occurrence rate of gas giants, which are massive enough to carve out deep gaps, so we search for a solution that creates a mass gradient in orbital period. 

We next introduce mass growth by envelope accretion, and we begin our discussion with dust-free atmospheres. As noted in Section \ref{subsubsec:cooling}, the cooling-limited accretion rate in dust-free atmospheres is sensitive to the disk temperature at the planet's location. Specifically, the runaway time scales like $\propto T_{\rm disk}^{3.75} M_{\rm core}^{-2.5}$. Outer planets therefore reach runaway faster, attain larger masses, and halt earlier by carving out deeper gaps.
This increasing-outward mass gradient plays a critical role in reversing the effect of gap-opening. For example, in the deep-gap limit, $t_{\rm mig} \propto M_p P^{0.1}$. To reproduce the slope of the observed occurrence rate $dN/d\log P \propto P \propto t_{\rm mig} \propto M_{\rm p} P^{0.1}$, $M_p \propto P^{0.9}$ is required.
 
\begin{figure*}[ht!]
\epsscale{1.2}
\plotone{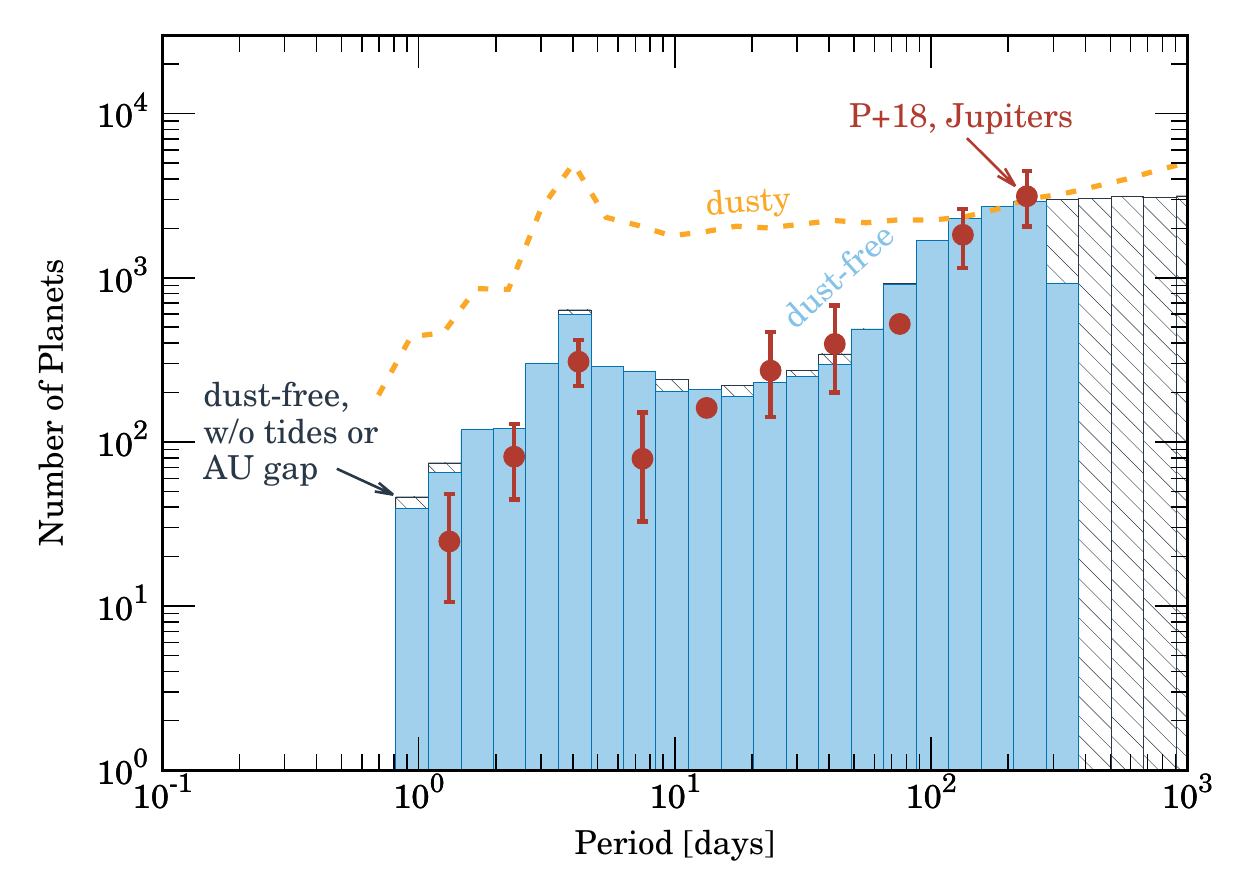}
\caption{Occurrence rates including envelope accretion and gap-opening for our fiducial model (in solid blue bars) and a model including dusty atmospheres (in orange). The measured occurrence rate for Jupiters is in red \citep{petmarwin18}. Our fiducial model consists of dust-free atmospheres and 18 $M_{\oplus}$ cores at disk depletion $d=0.055$. Cooling accretion in dust-free atmospheres becomes more efficient at larger orbital distances, meaning outer planets attain higher final masses, carve out deeper gaps, and slow their migration more rapidly than inner planets. Final planet masses in the fiducial setup cover $\sim$100 $M_{\oplus}$ at $\sim$10 days to $\sim$450 $M_{\oplus}$ at $\sim$100 days. The dusty planets begin with the same initial conditions, but reach equal planet masses ($\sim$34 $M_{\oplus}$) across all periods, because the cooling accretion rate in dusty atmospheres is insensitive to the nebular disk temperature. They reach a solution similar to that given in Figure \ref{figure:barecores} for equal-mass cores which carve out gaps during migration. Accounting for tidal interaction with the star or the photoevaporative gap at 1 AU alters the period distribution only at the edges of the dynamic range of our interest (compare the solid blue histogram with the dashed histogram, which omits these features). The Roche limit lies at $\approx$ 0.9 days.
\label{figure:fiducial}}
\end{figure*}

\begin{figure}
\centering
\includegraphics[width=0.5\textwidth]{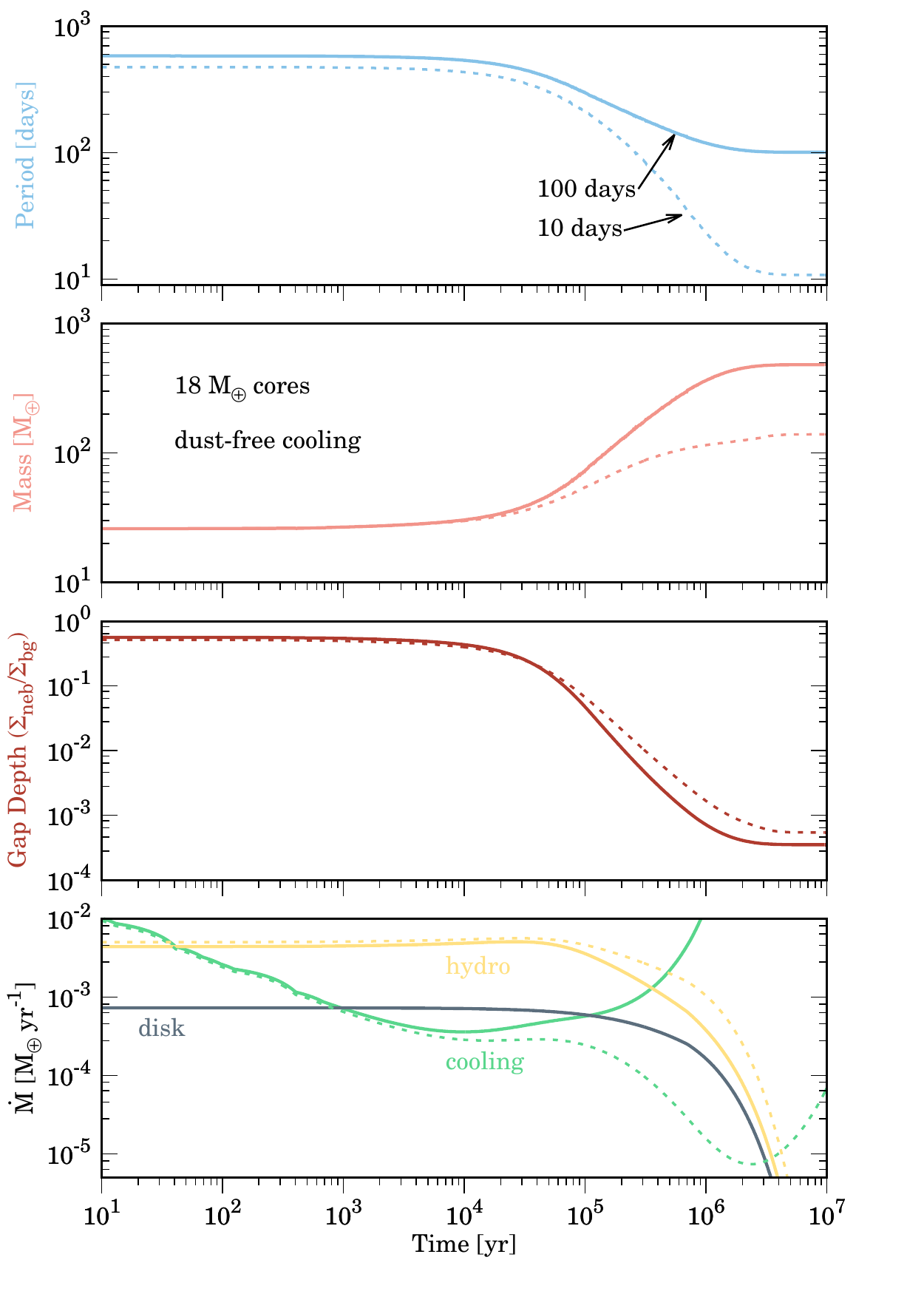}
\caption{The evolution of the planets which end at final periods of 10 and 100 days in our fiducial model (in dashed and solid curves, respectively). The accretion curves (bottom panel) show that the outer planet reaches runaway and its growth becomes disk-limited before the inner planet, which remains cooling-limited until beyond 1 Myr. The asymmetry in planet masses (second panel, light red) manifests itself in the gap depths ($(1+0.034K)^{-1}$, third panel, dark red); the outer, more massive planet clears out a deeper gap and slows down more rapidly than the inner planet. \label{figure:fiducial_evolution}}
\end{figure}

Figure \ref{figure:fiducial} illustrates in blue bars our fiducial model, in which we use 18 $M_{\oplus}$ cores, $d=0.055$, and dust-free atmospheres, which recovers the observed occurrence rates within 1-$\sigma$ for $P > 10$ days. We include both tidal dissipation and the 1 AU gap in our fiducial case. Distance-dependent mass growth and its effect on gap opening and therefore the migration rate are demonstrated in Figure \ref{figure:fiducial_evolution}. Growth in the mass of the inner planet is sluggish owing to the extended cooling-limited gas accretion, and the gap it carves out is relatively shallow, leading to faster migration. The outer planet triggers runaway accretion by $\sim$0.1 Myrs and its envelope growth becomes limited by global disk accretion. With a deeper gap, it slows down and lingers.

The orange curve in Figure \ref{figure:fiducial} is the result with the same initial conditions but assuming dusty atmospheres. The cooling-limited accretion rate for these models is insensitive to the planet's location in the disk. Inner and outer planets therefore attain similar final masses and so the occurrence rate profile flattens by gap opening. We note that with this setup, dusty accretion cannot produce Jupiter-mass planets. If we tune the parameters ($d\gtrsim 0.5$ and core masses $\gtrsim 22 M_{\oplus}$) so as to create gas giants, the result is the same---a vanishing mass gradient and a flat occurrence rate profile.

We observe no significant effect of accounting for tidal interaction between the host star and the planets. Only planets inside $\sim 1$ day by the end of the 10 Myr migration simulation spiral inward past the Roche limit at $0.9$ days where they are presumably destroyed and so are dropped from our calculation. Planets beyond this shuffle around, moving both inwards and outwards according to the initial orbital period and the spin-down evolution of the host star (Section \ref{subsec:tides}), but its effect is minimal and relevant only to the innermost extremities of the occurrence rate profile. We tested the effect of including tidal dissipation in the planet with non-zero initial eccentricity. Eccentricity tides boost the orbital decay rate out to $\sim$5 days, shifting the hot Jupiter peak from $\sim$4 days to $\sim$3 days and removing planets inside $\sim$2 days. Ultimately, for our chosen tidal dissipation constant ($\sigma_{\rm p}=2\cdot 10^{-60}~\rm{g}^{-1}~\rm{cm}^{-2}~\rm{s}^{-1}$;  \citealt{han10}) and initial eccentricity $\sim$0.1, the model hot Jupiter peak is too sharp compared to the data. We note that the 1 AU gap affects only the planets nearest the gap, depleting the occurrence there as planets are torqued outwards.

In Section \ref{subsec:migration}, we discussed how the minimum initial periods of planets were chosen to avoid an excess pile-up of planets at small periods. The weak tidal decay means that tidal effects are not capable of offering an alternative avenue of avoiding pile-up, where tides may redistribute planets to smaller or larger periods. This holds true for tidal dissipation constants up to $\sim$2 orders of magnitude larger than our fiducial value (roughly equivalent to tidal quality factors $Q_{\star}^{\prime} \simeq 10^{6}$--$10^8$), which only yield destruction of planets out to $\sim$2 days.

\begin{figure}
\centering
\includegraphics[width=0.5\textwidth]{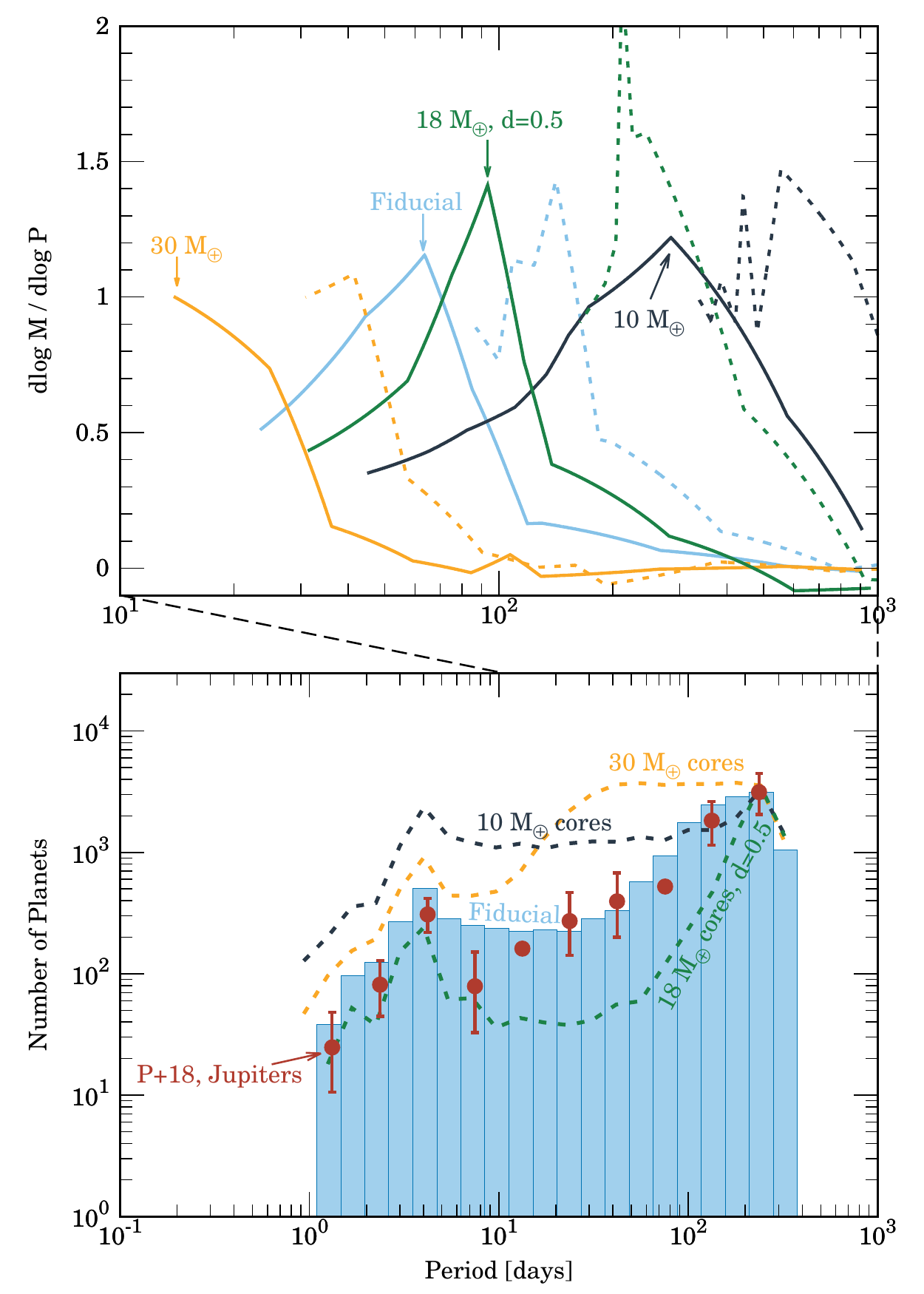}
\caption{Top: Planet mass gradients ($d\log M_{\rm p} / d\log P$) for various model parameters during active inward migration, at 1 Myr (solid curves) and 0.3 Myr (dashed curves). These models are computed at $d=0.055$, for 10 $M_{\oplus}$ (dark navy), our fiducial 18 $M_{\oplus}$ (in light blue, the same as in Figure \ref{figure:fiducial}), and 30 $M_{\oplus}$ cores (in orange). We also show the result with 18 $M_{\oplus}$ cores at $d=0.5$ (in green). Bottom: The corresponding occurrence rates for our example models overlaid on the measured rates from \citet[][red circles]{petmarwin18}. The top panel zooms into the range of orbital periods where we observe a rise in occurrence rates. Where mass gradients flatten, the occurrence rate profile flattens; where mass gradients rise, the occurrence rate profile steepens. See the main text in Section \ref{sec:masses} for more details on what brings about these mass gradients.
\label{fig:occurrence_compare}}
\end{figure}

\subsection{Disk Viscosity Parameter and the Global Accretion Rate}

The results for our fiducial disk parameters show that recovering the observed occurrence rate depends on the mass gradient of the model planet ensemble, which in turn is governed by the onset of cooling-limited accretion and runaway (for dust-free atmospheres; dusty atmospheres grow at a rate independent of orbital period and therefore produce a flat occurrence rate profile). Whether and when the mass growth switches between cooling-limited to hydrodynamically-limited is dependent on the global disk accretion rate $\dot{M}_{\rm disk}$.

Physically, changes in $\alpha$ and/or the gas depletion factor $d$ adjust $\dot{M}_{\rm disk}$.  Lower $\dot{M}_{\rm disk}$ flattens the occurrence rate profile as the mass differences between cooling-limited and disk-limited growth narrows. At $d=0.2$ and $\alpha=2.5\times 10^{-4}$, we recover the solutions obtained previously with $d=0.055$ from Figure \ref{fig:occurrence_compare}, since $\dot{M}_{\rm disk}$ is equivalent. Pushing to yet larger $d$ (i.e., larger $\dot{M}_{\rm disk}$) we find steeper occurrence profiles as mass gradients increase, akin to the $d=0.5$ solution in Figure \ref{fig:occurrence_compare}.

Adopting an actively accreting disk as treated in the radiative equilibrium models of \cite{dalcancal98}, we find that the observed occurrence profile can only be recovered with extreme fine-tuning of initial conditions. Their choice of $\alpha=10^{-2}$ yields enhanced corotation torques \citep{kantanszu18} which further produce a flattening of the model occurrence rates. The active disk model hinders our ability to recover the observations.

\subsection{Core Masses and Disk Masses}
\label{sec:masses}

The data can only be recovered within their errors in a limited region of parameter space, spanning roughly $M_{\rm core}=$15--22 $M_{\oplus}$ and $d=$0.035--0.085, highlighting how sensitive migration theory is to the initial disk and planetary parameters. Figure \ref{fig:occurrence_compare} illustrates this sensitivity, demonstrating the range of possible shapes of model occurrence rate profiles.

Ultimately, the orbital period distribution of gas giants is shaped by the period-dependent rate of migration, which in turn is governed by the mass gradient of the gap-opening planets. Figure \ref{fig:occurrence_compare} demonstrates how the set of parameters that effect steep occurrence rate profiles are characterized by high mass gradient $d\log M_{\rm p} / d\log P$ during the time of active inward migration.

The final planet masses are more uniform for low mass cores (10$M_\oplus$) because the timescale to runaway is on the order of the disk lifetime and so the envelope growth is largely limited by gas thermodynamics. In dust-free accretion for a fixed core mass, $M_{\rm p} \propto M_{\rm env}/M_{\rm core} \propto T_{\rm disk}^{-1.5} \propto P^{0.43}$ (see equation \ref{equation:dustfree_gcr}) shallower than the required mass gradient 0.9.

On the opposite end of the spectrum, high mass cores ($30M_\oplus$) produce an occurrence profile that is steep inside $\sim$40 days and flat outside. Beyond $\sim$40 days, the runaway timescale is so short that gas growth for all the planets outside $\sim$40 days is limited by global disk accretion which unifies the planet masses. Inside $\sim$40 days, cooling-limited accretion takes hold, driving planet masses smaller with decreasing period. Simultaneously, these massive cores undergo initially rapid inward migration so that the mass differences between inner and the outer planet diverges early on, producing a steep occurrence rate profile.

Higher initial disk gas density steepens the occurrence rate. Using the same core mass as the fiducial model but increasing $d$ to 0.5, we find the inner and outer planets evolve akin to Figure \ref{figure:fiducial_evolution}, except after runaway the outer planet grows in mass at the rate of global disk accretion, heightened by $d$. The inner planet grows at the cooling rate, which is boosted only by factors of $d^{0.12}$. The masses of the inner and the outer planets diverge more dramatically and we observe large $d\log M_{\rm p}/d\log P$.

\subsection{Comparing to the Giant Planet Mass Distribution}

In reality, initial core masses will not be uniform throughout the disk, and the disk mass will vary between many systems \citep[e.g.][]{androskra13}. 
Cores can build up to different masses at different orbital distances (e.g., pebble isolation mass \citealt{bitmorjoh18}); furthermore, the ability to reach those masses can be variable with respect to disk properties. Since the behavior of such variability is not well understood, we simulate it by sampling the core masses from a lognormal distribution centered at either a fixed mass or the pebble isolation mass using equations 10 and 11 of \citet{bitmorjoh18}.We further vary disk masses, sampling $d$ uniformly in a logarithmic sense, from $\log d = -3$ to $\log d = -0.3$. Sampling disks uniformly in a linear sense was found to over-produce the number of hot Jupiters beyond the modest bump observed; by sampling larger disk masses, migration times are shortened and too many planets are shuttled to inner disk edges. The minimum initial orbital distance we sample is 0.5 AU in all models.

\begin{figure*}
\centering
\includegraphics[width=1.\textwidth]{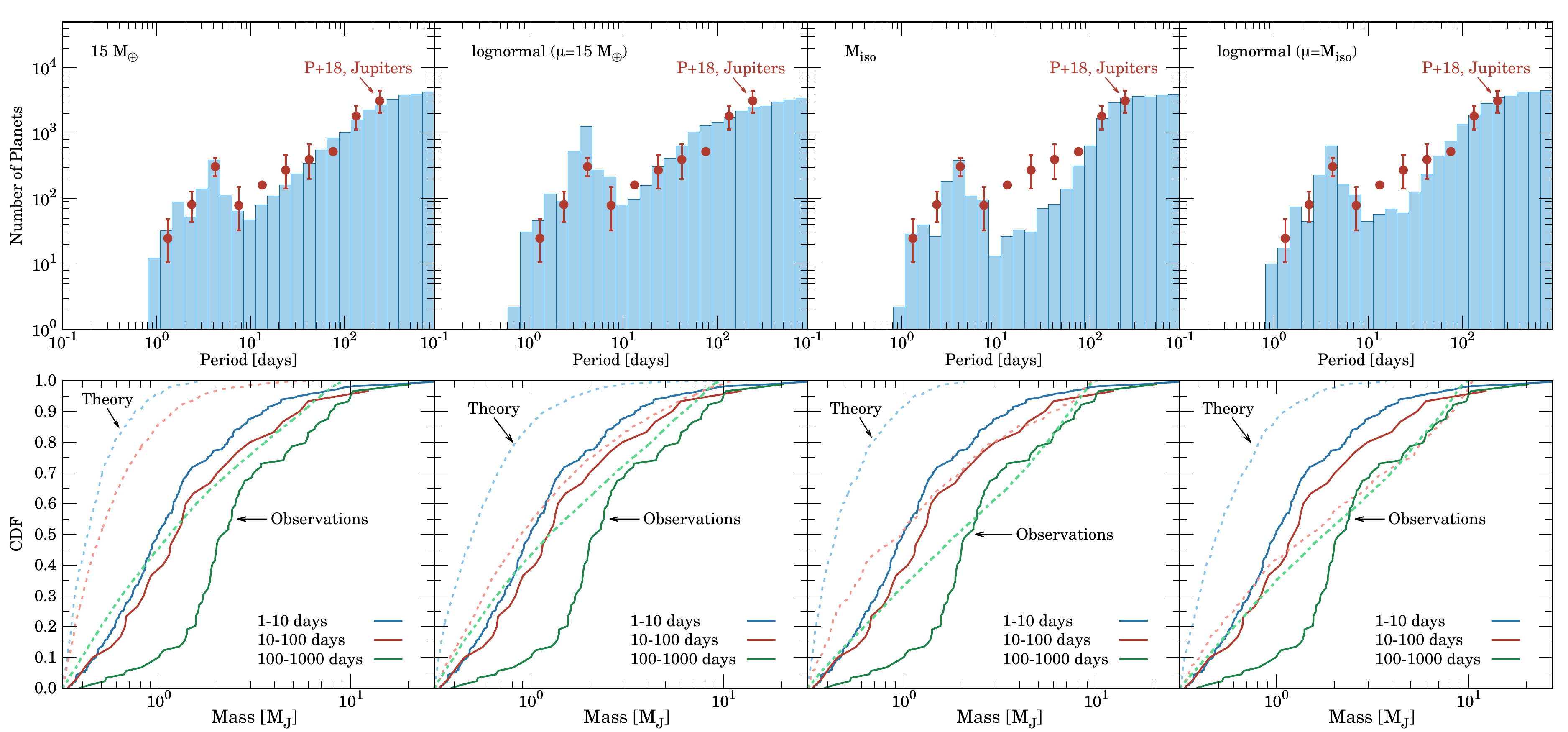}
\caption{Model occurrence rates (top row) and associated planet mass cumulative distribution functions (CDFs; bottom row) for various migration models. All models sample disk masses uniformly in log from $d=0.001$ to $d=0.5$. Models from left to right set core masses to a constant 15$M_{\oplus}$, sample from a lognormal distribution centered on 15$M_{\oplus}$ with standard deviation of 0.45, set to the pebble isolation mass \citep[e.g.][]{bitmorjoh18}, and sample from a lognormal distribution centered on the pebble isolation mass with standard deviation of 0.45. CDFs are broken into bins spanning 1-10 days (blue curves), 10-100 days (red curves), and 100-1000 days (green curves), with data in solid curves and theoretical predictions in dashed. Occurrence rate data in red points are from \cite{petmarwin18}, and the observed low-eccentricity ($e<0.1$) giant planet mass distribution is from the Exoplanet Archive (downloaded on Septemember 29, 2020; \url{https://exoplanetarchive.ipac.caltech.edu}). Only lognormal distributions centered near $\sim 12-18 M_{\oplus}$ recover both the occurrence profile and the warm Jupiter mass distribution. 
The bottom-heavy mass distribution for cold Jupiters may in principle be compatible with observations if biases against low-mass Jupiters are accounted for. The bottom-heavy distribution of hot Jupiter masses is common across all models and may be too severe even after correcting for biases.
\label{fig:final_masses}}
\end{figure*}

In Figure \ref{fig:final_masses} we illustrate the resulting occurrence profiles and planet mass cumulative distribution functions (CDFs) with disk masses drawn from a uniform-in-log distribution as described previously. Also plotted is the mass CDF for observed giant planets ($>=100 M_{\oplus}$) on low-eccentricity orbits ($e<0.1$). Setting the core mass to a constant $15 M_{\oplus}$ recovers the occurrence profile, but severely under-predicts the masses of both hot and warm Jupiters (see the first column in Figure \ref{fig:final_masses}). Increasing the core mass shifts the occurrence profile outside 1$\sigma$ to resemble the $30 M_{\oplus}$ profile in Figure \ref{fig:occurrence_compare}.

Sampling $M_{\rm core}$ from a lognormal distribution centered on 15 $M_{\oplus}$ with a standard deviation (in log space) 0.45 recovers the occurrence profile and moves the warm Jupiter CDF closer in agreement with the data (see the second column of Figure \ref{fig:final_masses}; we cannot reject the null hypothesis that they are drawn from the same distribution, with KS statistic 0.19, and p value 0.2). Hot Jupiter masses remain too low, and the order-unity warm to hot Jupiter number fraction is an order of magnitude lower than that observed. The bottom-heavy cold Jupiter mass distribution may still be compatible with observations considering observational biases. A broader distribution with standard deviation of 0.67 shifts the occurrence outside 1$\sigma$ error of the observed data and produces a warm Jupiter mass distribution that is too top-heavy compared to observations. Similar models centered on 30$M_{\oplus}$ again resemble the fixed $30 M_{\oplus}$ profile in Figure \ref{fig:occurrence_compare}, with top-heavy warm and cold Jupiter mass distributions well beyond that observed.

Setting core masses to the pebble isolation mass produces an occurrence profile too steep (third column in Figure \ref{fig:final_masses}). The mass CDF shows a severely bottom-heavy distribution of Hot Jupiter masses, and a nearly uniform mass CDF for planets beyond 100 days. The model CDF of warm Jupiter masses features what appears to be an overpopulation of $\lesssim 500 M_{\oplus}$ compared to the observations, though this discrepancy may be explained by observational bias against low-mass objects.

Sampling core masses from a lognormal distribution centered on the pebble isolation mass (i.e., the mean of the mass distribution increases with orbital distance) with the standard deviation of 0.45 brings the occurrence rate closer in agreement with observations, but produces a mass CDF for warm Jupiters which is too top-heavy (see the last column in Figure \ref{fig:final_masses}). A top-heavy warm Jupiter mass distribution cannot be reconciled with observational biases. Decreasing the standard deviation down to 0.15 brings the mass CDF closer to the isolation mass-only model, but the occurrence rates drop well below the observations. The pebble isolation mass models fail to simultaneously recover the mass spread and reasonable $dN/d\log P$ profiles.

\section{discussion and conclusion}\label{sec:discussion}

Our results underline how sensitive migration theory is to the disk and planet initial conditions. Although with an appropriate choice of both we are able to recover simultaneously the observed occurrence profile within 1-$\sigma$ and the mass distribution of warm Jupiters, the hot Jupiter mass distribution is always bottom-heavy, and warm to hot Jupiter number fractions $\sim 1$ are too low. We require distance-dependent opacities (as provided by dust-free atmospheres), and core masses distributed lognormally centered near $\sim 12-18 M_{\oplus}$, with a non-zero spread less than 0.67. Figure \ref{fig:final_masses} places these results in context with the observed masses of confirmed, low-eccentricity Jupiters.

Previous work on generating orbital period distributions of gas giants invoked disk gas dispersal \citep[e.g.,][]{trilunben02,armlivlub02}, photoevaporative gap opening \citep[e.g.,][]{alepas12}, and disks with planet traps caused by varying opacities \citep[e.g.,][]{colnel16,alidib17}. Our model incorporates many of the same physics (with the exception of planet traps intrinsic to disk structures). By dissecting each process and directly comparing the model result to the observed occurrence rate of gas giants, we identify the importance of the interplay between gas accretion onto cores and gap-opening by the planet in setting the period distribution of giant planets.

In the migration paradigm, the overall rise of the gas giant occurrence rate requires planets to move in more rapidly close to the star and stall farther out. For a given disk profile, the rate of migration is ultimately determined by the mass of the planet so that there is a correlation between the slope of the number of gas giants per star vs.~period and the slope of the final mass of gas giants vs.~period. Only a narrow range of parameter space reconciles both the observed occurrence rate and mass distribution of warm Jupiters, and none of our models can account for high-mass hot Jupiters. The large-scale migration paradigm does not appear to robustly predict the giant planet occurrence rate concurrently with their mass distribution, disfavoring disk-induced migration as the dominant origin of warm and hot Jupiters. Given the observational biases inherent in the observed mass distribution, we note that it is difficult to say definitively that the bottom-heavy mass distributions are truly incompatible with the data.

\subsection{Caveats}\label{sec:caveats}
Our model assumed the most basic form of disk structure so that it can be easily generalized.  Details of three-dimensional hydrodynamic, thermodynamic, and magnetohydrodynamic processes can impact the migratory behavior.

Departure from locally isothermal equations of state may pose corrections to migration theory. This has been shown to enhance the role of the corotation torque \citep{paabarcri10}, and alter disk angular momentum flux in spiral waves \citep{mirraf20,ziakledul20}. \cite{kantanszu18} note that for planets with shallow gaps the enhanced corotation torque could significantly slow migration rates. \cite{paabarkle11} further show that thermal diffusion can sustain corotation torques for low-mass planets. Diffusive effects may therefore also slow migration before our planets carve out deep gaps. Most of our model planets readily carve out deep gaps where the deficit of material in the horseshoe region downplays the magnitude of corotation torques. Deviations from local isothermality and thermal diffusion may only have a minor effect on our results.

There are many ways in which migration can be halted or reversed.
Radially varying opacities can create radial substructures in the disk acting as planet traps \citep[e.g.,][]{masmorcri06,colnel16}. Outward torques from the heat of solid accretion \citep{benmasskoe15}, or resonant capture of companions (e.g., \cite{raymansig06}; see \cite{huawutri16} for observational evidence of close, small companions to warm Jupiters) can also complicate the picture. The presence of multiple, close-in giant companions could in principle alter our period distributions if configurations become unstable after disk dispersal. This may partially explain the warm Jupiter eccentricity distribution \citep{andlaipu20}. However, most gas giants inside 1 AU are either alone or neighbored by small planets.

Our treatment of the photoevaporative gap at $\sim$1 AU assumes the gap becomes wide enough, quickly enough, such that the inner torque dominates the outer torque. While our choice is motivated by Figure 9 of \citet{oweerccla11}, the presence of a planet is known to trigger more rapid drain-out of the inner disk \citep[e.g.,][]{rosercowe13} and may be another means of stalling migration. A more careful time-dependent, grid-based prescription of disk photoevaporation that accounts for feedback between envelope accretion, migration, and photoevaporation would be useful. 

An important caveat to our results is that we assumed steady-state viscous disks. The clean rings of dust emissions seen in ALMA images suggest protoplanetary disks to be close to inviscid ($\alpha < 10^{-4}$; e.g., \citealt{pindenmen16,flahugros17}).  Without `viscous' diffusion to fill planetary gaps back in, planets in inviscid environments may halt their migration rapidly \citep[e.g.,][]{Li09,Yu10,funchi17}. Vortices may also introduce episodic inward radial forcing for giant planets \citep{mcnnelpaa19,mcnnelpaa20}. \cite{idatanjoh18} propose that migration may also slow in laminar MHD disks if the $\alpha$ for wind-driven accretion outweighs a separate, gap-opening $\alpha$. Gas accretion in inviscid disks also differs from our treatment. Gas may be supplied by magnetized disk winds \citep{baisto13,greturnel15,wangoo17,roschigin20}, repulsive torques from companion planets \citep{gooraf01,sargol04,funchi17}, or the Hall effect \citep{leskunfro14,simleskun15,bai15}. 

\subsection{Implications for the Migration Paradigm}

The potential formation pathways of warm and hot Jupiters can be divided into three categories: in situ formation, high-eccentricity migration, and disk-induced migration. Given their wide range of eccentricities, warm Jupiters likely form through at least two pathways - one to account for the circular population (the in-situ or disk migration pictures; e.g., \citealt{colnel16}), and one for the eccentric population (the high eccentricity migration or scattering pictures; e.g., \citealt{jurtre08,dawchi14,pettre16}). Given that high-eccentricity migration is not sufficient to explain most warm Jupiters, we are thus left with disk migration and in situ formation.

The period distribution's sensitivity to the planet and disk initial conditions in the {\it large-scale} migration paradigm, along with the migration picture's difficulty in simultaneously accounting for the distribution of masses and periods of giant planets, challenges the hypothesis that large-scale migration is the dominant physical process setting the giant planet occurrence rate. Given the multitude of ways migration may be slowed and even halted, we propose instead that in situ assembly of warm Jupiters may provide a more robust explanation. Initial epochs of migration may well be inevitable, but the picture in which giant planets arrive at their final periods and masses principally via long-range migration through the disk appears inconsistent with their occurrence rate and mass distribution. Whether in situ formation of giant planets can quantitatively explain the observed occurrence rate profile is the subject of our future work. We also advocate for future work characterizing the unbiased giant planet mass distribution to provide a cleaner and more powerful test of theoretical predictions.

\acknowledgments

We thank the anonymous referee for their prompt review that helped make the manuscript better. We would also like to thank Heather Knutson for insightful discussions. T.H. was supported in part by the Natural Sciences and Engineering Research Council of Canada (NSERC) through an Alexander Graham Bell Scholarship and by a TEPS-CREATE fellowship. E.J.L. is supported by the NSERC, RGPIN-2020-07045. This research was enabled in part by support provided by Calcul Qu\'{e}bec (\url{www.calculquebec.ca}), Compute Ontario  (\url{computeontario.ca}) and Compute Canada (\url{www.computecanada.ca}).
Figures were produced using \texttt{gnuplot}.

\bibliography{hallatt}{}
\bibliographystyle{aasjournal}

\end{document}